\documentstyle[amstex,amssymb,aps,twocolumn,floats,epsfig]{revtex}
\begin{document}
\title{Nonlocal mixing of supercurrents in Josephson ballistic point contact}
\author{Yu.A. Kolesnichenko, A.N. Omelyanchouk, S. N. Shevchenko}
\address{B.I. Verkin Institute for Low Temperature Physics and Engineering,\\
National\\
Academy of Sciences of Ukraine, 47 Lenin Ave., 61103, Kharkov,\\
Ukraine}
\date{\today}
\maketitle
\begin{abstract}
We study coherent current states in the mesoscopic superconducting weak
link simultaneously subjected to the order parameter phase difference $%
\phi $ on the contact and to the tangential to the junction interface
superfluid velocity ${\bf v}_{s}$ in the banks. The Josephson current-phase
relation $I_{J}(\phi )$ controlled by the external transport current ${I}%
_{T}({\bf v}_{s})$ is obtained. At $\phi $ close to $\pi $ the nonlocal
nature of the Josephson phase-dependent current results in the appearance of
two vortexlike states in the vicinity of the contact.
\end{abstract}


\vspace{-1cm}

The superfluid flow of Cooper pairs in superconductor is related
to the space dependence of the phase $\chi $ of the order
parameter. In (quasi)homogeneous current state the supercurrent
density ${\bf j}$ locally depends on the superfluid velocity ${\bf
v}_{s}={\frac {\hbar}{2m}} \nabla \chi ({\bf r})$. Such a state is
realized in narrow films or wires \cite{deGen}. In the case when
the phase $\chi $ strongly varies in the scale of superconducting
coherence length $\xi _{0}$ the relation between the current density ${\bf j}%
({\bf r})$ and $\chi ({\bf r})$ becomes nonlocal. This situation (opposite
to the homogeneous current state) is realized in Josephson weak links (for
review see \cite{Likharev}), e.g. in superconducting point contacts -
microconstrictions between two bulk superconductors (banks). The Josephson
current nonlocally depends on $\chi ({\bf r})$ and is determined
(parameterized) by the total phase difference $\phi $ across the weak link.
The current-phase relation ${\bf j}(\phi )$ for ballistic point contact was
obtained in \cite{KO}. The nonlocal nature of Josephson current in
mesoscopic junctions was demonstrated by Heida {\it et. al.} \cite{Heida}
and studied in theoretical papers \cite{ZBO}.

The Josephson weak link could be considered as a "mixer" of two
superconducting macroscopic quantum states in the banks. The result of the
mixing is the phase dependent current carrying state with current flowing
from one bank to another. The properties of this state depend on the
properties of the states of the banks. For example, in Josephson junction
between unconventional ($d$-wave) superconductors the surface current,
tangential to the contact interface, appears simultaneously with Josephson
current (see {\it e.g.} \cite{AOZ}).

In this paper we study coherent current states in the Josephson
weak link between conventional superconductors, whose banks are in
the homogeneous current states. The questions raised in our
consideration are the following ones: How two superconducting
current carrying states in the banks are coherently mixed by a
mesoscopic Josephson junction, or in other words, what is a result
of the interplay between transport current{\bf \ }$j_{T}({\bf
v}_{s})$
flowing parallel to the junction interface and nonlocal Josephson current $%
j_{J}(\phi )$? How the Josephson properties of the system are
influenced by the external controlling transport current? We have
found that the distribution of the current in a region of nonlocal
mixing strongly depends on the global phase difference $\phi $
between banks and for $\phi =\pi $ contains the vortexlike states.
The current-phase relation $j_{J}(\phi )$ at $\phi $ near $\pi $
essentially depends on the superfluid velocity in the
banks ${\bf v}_{s}$, in particular, the absolute value of the derivative $%
dj_{J}/d\phi $ at $\phi =\pi $ is suppressed by the transport current.

We consider the Josephson weak link with direct conductivity - a microbridge
between thin superconducting films. The bridge sizes, length $L$ and width $%
2a$, are assumed to be smaller than the coherence length $\xi _{0}$. In this
case even for temperature $T$ near the critical temperature $T_{c}$ the
local description based on the Ginzburg-Landau approach is not applicable.
To describe the coherent current states in the system we use the
quasiclassical Eilenberger equations\cite{Eilen}, which are valid for
temperatures $0<T<T_{c}$ and for arbitrary relation between the contact size
and coherence length $\xi _{0}$. On the other hand, we assume that $a$\ and $%
L$ are much larger than the Fermi wavelength $\lambda _{F}$. The electron
mean free path is supposed to be much larger than $\xi _{0}$.

Suppose the homogeneous transport current $I_{T}$ with a superfluid velocity
${\bf v}_{s}$ flows in the banks of the contact. The situation with
controlled phase difference $\phi $ and preset current $I_{T}$ may be
realized if the microbridge is incorporated in a cylindrical thin film (Fig
1). Let the radius of the cylinder be less than London penetration depth and
larger than the coherence length. In this case the phase difference $\phi $
is governed by the external magnetic flux $\Phi ,$ $\phi =\frac{2e}{\hbar c}%
\Phi ,$ and the external transport current $I_{T}$ flowing along the
cylinder is homogeneously distributed far from the microconstriction.

\begin{figure}
\vspace{0mm}
\begin{center}
\leavevmode
    \epsfxsize=40mm
 \epsfbox{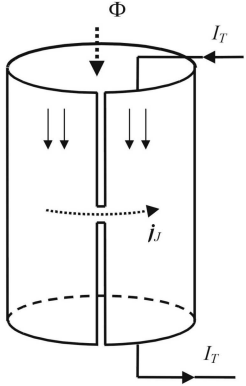}
 \end{center}
\vspace{0in} \caption{{\protect\small Scheme of the realization of
the microbridge with transport current in the banks and controlled
phase difference. }}
\end{figure}

The Eilenberger equations for the $\xi $-integrated Green's
functions have the form \cite{Eilen}:
\begin{equation}
{\bf v}_{F}\frac{\partial }{\partial {\bf r}}\widehat{G}_{\omega }({\bf v}%
_{F},{\bf r})+[\omega \widehat{\tau }_{3}+\widehat{\Delta }({\bf v}_{F},{\bf %
r}),\widehat{G}_{\omega }({\bf v}_{F},{\bf r})]=0,  \label{EqA1}
\end{equation}
where
\begin{equation}
\widehat{\Delta }=\left(
\begin{array}{cc}
0 & \Delta \\
\Delta ^{\dagger } & 0
\end{array}
\right) ,\quad \widehat{G}_{\omega }({\bf v}_{F},{\bf r})=\left(
\begin{array}{cc}
g_{\omega } & f_{\omega } \\
f_{\omega }^{\dagger } & -g_{\omega }
\end{array}
\right) ;
\end{equation}
$\widehat{\tau }_{3}$ is the Pauli matrix, $\Delta $ is the superconducting
order parameter, and $\widehat{G}_{\omega }({\bf v}_{F},{\bf r})$ is the
matrix Green's function, which depends on the electron velocity on the Fermi
surface ${\bf v}_{F}$, the coordinate ${\bf r}$, and the Matsubara frequency
$\omega =(2n+1)\pi T$, with $n$ being an integer number.

The order parameter $\Delta $ is determined by the self-consistency equation

\begin{equation}
\Delta ({\bf r})=\pi \lambda T\sum\limits_{\omega }\left\langle f_{\omega }(%
{\bf v}_{F},{\bf r})\right\rangle _{{\bf v}_{F}}.  \label{GapEq}
\end{equation}
Solution of the matrix equation (\ref{EqA1}) together with
Eq.(\ref{GapEq}) determines the current density ${\bf j(r)}$ in
the system

\begin{equation}
{\bf j(r)}=-2\pi ieN(0)T\sum\limits_{\omega }\left\langle {\bf v}%
_{F}g_{\omega }({\bf v}_{F},{\bf r})\right\rangle _{{\bf v}_{F}}.
\label{current}
\end{equation}
Here $\lambda $ is the BCS coupling constant, $N(0)$ is the density of
states at the Fermi surface, $\left\langle ...\right\rangle _{{\bf v}_{F}}$
is the averaging over directions of the velocity ${\bf v}_{F}$.

If the film thickness $w$ is much smaller than $\xi _{0},$ the
spatial distributions of $\Delta ({\bf r})$ and ${\bf j(r)}$
depend only on coordinates in the plane of the film and the
Eilenberger equations (\ref{EqA1}) reduce to the two-dimensional
ones. We solve these equations in the model of the microbridge as
a slit in thin impenetrable partition ($L=0$) at $x=0$ between two
half-planes $x\lessgtr 0$ (Fig. 2). The equations (\ref{EqA1}) for
Green's function $\widehat{G}_{\omega }({\bf v}_{F},x,y)$ have to
be supplemented by the continuity condition at the slit $\left(
x=0,\left| y\right| <a\right) $ and by the condition of the
specular reflection at the line $(x=0,\left| y\right| \geqslant
a)$. Far from the constriction the Green's functions must satisfy
to the conditions, which describe the homogeneous current parallel
to the $y-$axis.

\begin{figure}
\vspace{0mm}
\begin{center}
\leavevmode
    \epsfxsize=60mm
 \epsfbox{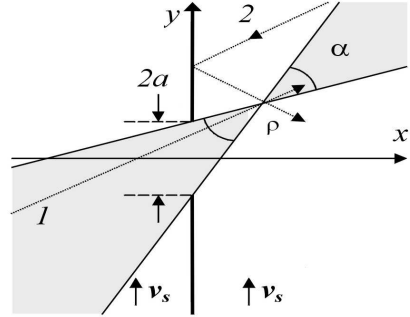}
 \end{center}
\vspace{0in} \caption{{\protect\small Model of the microbridge as
the slit in the thin insulating partition }} \label{1ris}
\end{figure}

As it was shown in \cite{KO} in the zero approximation on the small
parameter $a/\xi _{0}\ll 1$ for a self-consistent calculation of the
superconducting current it is not necessary to find the spatial dependence $%
\Delta (x,y)$. In the same approximation the superfluid velocity ${\bf v}%
_{s} $ does not depend on coordinates. The spatial variation of $\Delta $
and ${\bf v}_{s}$ is essential at the distances $\rho $ from the contact $%
\rho \lesssim a$. The Green's functions are varied at the distances of order
$\xi _{0}$ and in the main approximation on the parameter $a/\xi _{0}\ll 1$
they are defined by the values of $\Delta \left( {\bf v}_{s}\right) $ and $%
{\bf v}_{s}$ in the banks of the contact. For $\Delta $ and ${\bf v}_{s}$
being constants at each half-plane an analytical solution of Eilenberger
equations can be found by the method of integration along quasiclassical
trajectories. Under the condition $a/\xi _{0}\ll 1$ such solution is
selfconsistent. In any point ${\bf \rho =}\left( x,y\right) $ all ballistic
trajectories can be classified as transit trajectories (marked by ''$1$''\
at Fig.2), for which ${\bf v}_{F}$ $\in \alpha ({\bf \rho })$ ($\alpha ({\bf %
\rho })$ being the angle at which the slit is seen from the point ${\bf \rho
}$) and non-transit trajectories, ${\bf v}_{F}$ $\notin \alpha ({\bf \rho })$%
, (marked by ''$2$''\ at Fig. 2). For transit trajectories the Green's
functions satisfy the boundary conditions in the both banks. For the
non-transit trajectories they satisfy to specular reflection condition at
the partition and the conditions in the left or right bank. Making use of
the solution of Eilenberger equations, we obtain the following expression
for the current density (\ref{current}) at the slit:

\begin{gather}
{\bf j}(x=0,\left\vert y\right\vert <a,\phi ,{\bf v}_{s})=  \label{j(0)} \\
4\pi \left\vert e\right\vert N(0)v_{F}T\sum\limits_{\omega >0}\left\langle
\widehat{{\bf v}}Im\frac{i\Omega \sin \frac{\phi }{2}-\eta \widetilde{\omega
}\cos \frac{\phi }{2}}{\eta \Omega \cos \frac{\phi }{2}-i\widetilde{\omega }%
\sin \frac{\phi }{2}}\right\rangle _{\widehat{{\bf v}}},  \nonumber
\end{gather}
where $\Omega =\sqrt{\widetilde{\omega }^{2}+\Delta ^{2}}$, $\widetilde{%
\omega }=\omega +i{\bf p}_{F}{\bf v}_{s}$, $\widehat{{\bf v}}={\bf v}%
_{F}/v_{F}$ is the unit vector, $\eta =sign(v_{x})$. We should require $%
Re\Omega >0,$ which fixes the sign of the square root to be $sign({\bf p}_{F}%
{\bf v}_{s})$. Under the condition $a/\xi _{0}\ll 1$ the current density at
the slit does not depend on the $y-$coordinate, and the total current
through the contact (Josephson current) is equal to $I_{J}=2awj_{x}(x=0,%
\left\vert y\right\vert <a,\phi ,{\bf v}_{s})$.

For ${\bf v}_{s}=0$ the component of the current (\ref{j(0)}) tangential to
the contact $j_{y}\equiv 0$ and for the Josephson current density $%
j_{J}\equiv j_{x}$ we have the result obtained in the paper \cite{KO}. In
general case ${\bf v}_{s}\neq 0$ the current (\ref{j(0)}) has both $j_{J}$
and $j_{y}$ components. The tangential current $j_{y}$ depends on the phase $%
\phi $ and is not equal to the transport current density ${\bf
j}_{T}$ in the banks. In particular, at $\phi $ near $\pi $ it
goes in opposite direction to the external transport current (see
below).

To describe the influence of the transport current in the banks on the
Josephson current we introduce the dimensionless parameter $%
q=v_{s}p_{F}/\Delta _{0}$ $\ \left( \Delta _{0}=\Delta
(T=0,v_{s}=0)\right).
$ The value of $q$ is varied in the range $0<q<q_{c}$. The critical value $%
q_{c}$ corresponds to the critical current density in the
homogeneous current state. At zero temperature $q_{c}=1$, and the
gap $\Delta $ does
not depend on $q$ \cite{Bardeen}. In Fig.3 we plot the Josephson current $%
I_{J}(\phi )$ at temperature $T=0.1T_{c}$ for different values of
$q$. The presence of the tangential transport current in the banks
suppresses the value of the critical Josephson current and
essentially changes the derivative $dI_{J}/d\phi $\ at $\phi =\pi
$. We emphasize, that the
dependence of the Josephson current $I_{J}$\ $\left( \phi \right) $\ on $q,$%
\ which is shown in Fig.3, does\ not relate to the suppression of the gap by
the transport current, which is negligible for such low temperatures.

\begin{figure}
\vspace{0mm}
\begin{center}
\leavevmode
    \epsfxsize=70mm
 \epsfbox{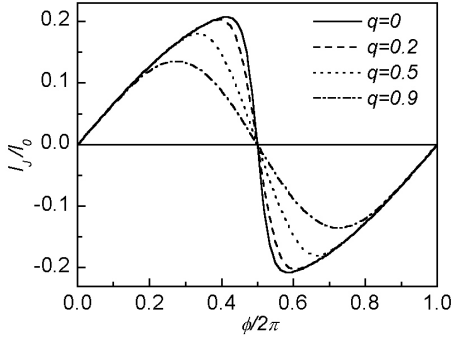}
 \end{center}
\vspace{0in} \caption{{\protect\small \ Josephson current $I_{J}$
versus phase $\protect\phi $. $I_{0}=2aw\cdot 4\protect\pi \left|
e\right| N(0)v_{F}T_{c}$.}}
\end{figure}

The derivative $dI_{J}/d\phi $\ at $\phi =\pi $\ determines the kinetic
inductance of Josephson junction \cite{Likharev}, which is relevant, e.g.,
for SQUID's operation. The expression for the derivative of the Josephson
current $I_{J}\ \ $at $\phi =\pi $\ has the form
\begin{equation}
\left. \frac{dI_{J}}{d\phi }\right| _{\phi =\pi }=-2aw\frac{|e|N(0)\Delta
^{2}}{mv_{s}}f\left( \frac{v_{s}p_{F}}{\pi T}\right) ,  \label{deriv}
\end{equation}
where function $f(x)$ is plotted in Fig.4.\ The derivative $dI_{J}/d\phi
(\phi =\pi )$\ is inversely proportional to $v_{s}$\ at $T\ll p_{F}v_{s}$\
and it is inversely proportional to $T$\ at $T\gg p_{F}v_{s}$.

\begin{figure}
\vspace{0mm}
\begin{center}
\leavevmode
    \epsfxsize=70mm
 \epsfbox{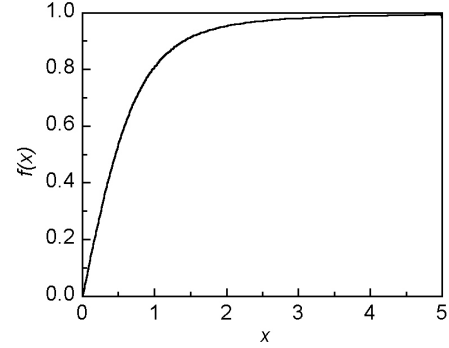}
 \end{center}
\vspace{0in} \caption{The function $f(x)$ from expression
(\ref{deriv})}.
\end{figure}

By using the Green's functions along transit and non-transit trajectories,
calculated in the main approximation on the small parameter $a/\xi _{0}$, we
can find the spatial distributions of the order parameter and the current
density in the contact (see Ref.\cite{KO}). The numerically calculated
current density distributions for different values of phase $\phi $ and the
temperature $T=0.1T_{c}$ are shown in Figures 5 and 6. For small values of
the phase difference $\phi $ between banks, the current density ${\bf j}(%
{\bf \rho })$ is just the vector sum of the homogeneous transport current
density in the banks ${{\bf j}_{T}}({{\bf v}_{s}})$ and the conventional
Josephson current ${\bf j}_{J}(\phi ,{\bf \rho ,v}_{s}=0)$ (Fig.5). For $%
\phi $ near $\pi $ the constructive interference of supercurrents takes
place. At $\phi =\pi $ there are no Josephson current, $j_{J}=0,$ and the
current is distributed in the way that there are two antisymmetric
''vortices''\ close to the contact region (Fig.6). Far from the constriction
(at the distances $\rho \sim \xi _{0}\gg a$) the interference current is
spread out and the current density is equal to its value in the banks.

\begin{figure}
\vspace{0mm}
\begin{center}
\leavevmode
    \epsfxsize=60mm
 \epsfbox{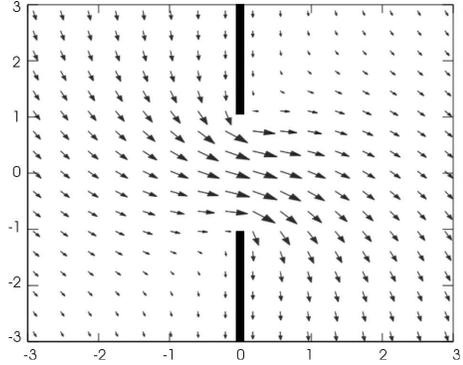}
 \end{center}
\vspace{0in}
\caption{{\protect\small Vector plot of the current density for $\protect\phi%
=\protect\pi/2$ and $q=0.5$. Numbers mark $x$ and $y$ axes in the
units of the contact size $a$.}}
\end{figure}

\begin{figure}
\vspace{0mm}
\begin{center}
\leavevmode
    \epsfxsize=60mm
 \epsfbox{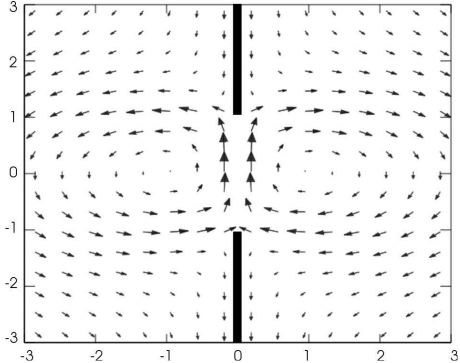}
 \end{center}
\vspace{0in} \caption{Vector plot of the current density for
$\protect\phi =\protect\pi $ and $q=0.5$. }
\end{figure}
Simple and transparent expression for the current density distribution ${\bf %
j(\rho )}$ can be found for temperatures close to the critical temperature ($%
T_{c}-T\ll T_{c}$). At the distances from the contact, which are less than
the coherence length $\xi _{0}$ and arbitrary in comparison with the size $a$%
, ${\bf j(\rho )}$ takes the form:
\begin{gather}
{\bf j(\rho ,}\phi {\bf ,v_{s})}={\bf j}_{J}({\bf \rho ,}\phi )+{\bf j}_{T}(%
{\bf v}_{s})+{\bf j}_{JT}({\bf \rho ,}\phi ,{\bf v}_{s});  \label{sum_J} \\
{\bf j}_{J}({\bf \rho ,}\phi )=2j_{c}\sin \phi \left\langle \widehat{{\bf v}}%
sign(v_{x})\right\rangle _{\widehat{{\bf v}}\in \alpha ({\bf \rho })};
\nonumber \\
{\bf j}_{T}({\bf v}_{s})=-j_{c}k\left\langle \widehat{{\bf v}}\widehat{v}%
_{y}\right\rangle _{\widehat{{\bf v}}};  \nonumber \\
{\bf j}_{JT}({\bf \rho ,}\phi ,{\bf v}_{s})=j_{c}k(1-\cos \phi )\left\langle
\widehat{{\bf v}}\widehat{v}_{y}\right\rangle _{\widehat{{\bf v}}\in \alpha (%
{\bf \rho )}},  \nonumber
\end{gather}
where
\begin{equation}
j_{c}\left( T,v_{s}\right) =\frac{\pi |e|N(0)v_{F}}{8}\frac{\Delta
^{2}\left( T,v_{s}\right) }{T_{c}}
\end{equation}
is a critical current density of the contact at $T\approx T_{c}$, $%
k=(14\varsigma (3)/\pi ^{3})(v_{s}p_{F}/T_{c})$. We detach explicitly the
Josephson current ${\bf j}_{J}({\bf \rho ,}\phi )$, and the spatially
homogeneous (transport) current density ${\bf j}_{T}({\bf v}_{s})$ produced
by the superfluid velocity ${\bf v}_{s}$, and write the total current (\ref
{sum_J}) as the sum of three components: ${\bf j}_{J}$, ${\bf j}_{T}$, and \
the rest - the ''interference''\ current ${\bf j}_{JT}$. The macroscopic
quantum interference takes place in the vicinity of the contact region where
both coherent current densities ${\bf j}_{J}({\bf \rho ,}\phi )$ and ${\bf j}%
_{T}({\bf v}_{s})$ exist. We emphasize, that at $\phi $ near $\pi $ at the
slit the ''interference''\ current ${\bf j}_{JT}$ is antiparallel to ${\bf j}%
_{T}.$ If the phase difference $\phi =\pi ,$ the current ${\bf j}_{JT}=-2%
{\bf j}_{T}$. When there is no phase difference (at $\phi =0)$, we have $%
{\bf j}_{JT}=0$.

In conclusion, we have investigated the coherent current states in
the Josephson ballistic point contact simultaneously subjected to
the order parameter phase difference $\phi $ and to the tangential
to the junction interface the superfluid velocity ${v}_{s}$ in the
banks. The current-phase relation $I_{J}(\phi )$ is shown to be
controlled by the transport
superconducting current $I_{T}({\bf v}_{s})$. Thus, varying $I_{T}({\bf v}%
_{s})$, the characteristics of the weak link, such as the shape of the
Josephson current-phase relation and the value of the critical current, can
be changed. The similar effect can be produced by the increasing of the
temperature $T$ of the system. But, as compared to the controlling by the
transport supercurrent, the increasing of the temperature leads to
additional thermal noise.

Moreover, the current distribution pattern in the vicinity of the
contact was obtained. The current pattern drastically depends on
the external phase difference $\phi $; in particular, at $\phi
=\pi $ the existence of two antisymmetric vortexlike current
structures is predicted. Considering the current pattern, we have
also demonstrated that the superposition of the supercurrents in
the vicinity of the weak link is not just their vector sum. These
results can be relevant in a wide range of problems, in which the
current (and corresponding magnetic field) distribution in the
vicinity of the weak link is important.

\bigskip We thank I.Dmitrenko, I.Yanson, A.Zagoskin and A.Amin for
stimulating discussions. We acknowledge support from D-Wave
Sys.Inc.(Vancouver).

\end{document}